\documentclass[prb,twocolumn,showpacs]{revtex4}
\usepackage{graphicx}
\begin{document}
\title{Simultaneous existence of two spin-wave modes
in ultrathin Fe/GaAs(001) films studied by Brillouin Light Scattering:
 experiment and theory}
\author{M. G. Pini$^{1,2}$}
\email{mgpini@ifac.cnr.it}
\author{P. Politi$^{1,2}$}
\author{A. Rettori$^{2,3}$}
\author{G. Carlotti$^{4,5}$}
\author{G. Gubbiotti$^{5}$}
\author{M. Madami$^{4,5}$}
\author{S. Tacchi$^{4,5}$}
\affiliation{
$^1$ Istituto di Fisica Applicata ``N. Carrara",
CNR, Via Madonna del Piano, I-50019 Sesto Fiorentino, Italy
\\
$^2$ INFM, Unit\`a di Firenze, Via G. Sansone 1, I-50019 Sesto Fiorentino, Italy
\\
$^3$ Dipartimento di Fisica, Universit\`a di Firenze,
Via G. Sansone 1, I-50019 Sesto Fiorentino, Italy
\\
$^4$ Dipartimento di Fisica, Universit\`a di Perugia,
Via Pascoli, I-06123 Perugia, Italy
\\
$^5$ INFM, Unit\`a di Perugia, Via Pascoli, I-06123 Perugia, Italy
}
\date{\today}
\begin{abstract}
A double-peaked structure was observed in the {\it in-situ} Brillouin
Light Scattering (BLS) spectra of a 6 \AA$~$ thick epitaxial
Fe/GaAs(001) film for values of an external magnetic field $H$, applied
along the hard in plane direction,
lower than a critical value $H_c\simeq 0.9$ kOe. This experimental finding is theoretically
interpreted in terms of a model which assumes a non-homogeneous magnetic ground state
characterized by the presence of perperpendicular up/down stripe domains.
For such a ground state, two spin-wave modes, namely an acoustic and an optic mode, can exist.
Upon increasing the field the magnetization tilts in the film plane,
and for $H \ge H_{c}$ the ground state is homogeneous,
thus allowing the  existence of just a single spin-wave
mode.
The frequencies of the two spin-wave modes
were calculated and successfully compared with the experimental data.
The field dependence of the intensities of the corresponding two peaks
that are present in the BLS spectra was also estimated, providing further
support to the above-mentioned interpretation.
\end{abstract}
\maketitle

\section{Introduction}

Ferromagnetic-semiconductor heterostructures, like ultrathin Fe/GaAs(001) films,
have received considerable attention\cite{Brockmann,Gester,Steinmueller,Monchesky}
for their potential technological applications in new magnetoelectronic
devices.\cite{Prinz,libro} A sharp and well ordered interface,
without any dead magnetic layer,  can be obtained, as bcc Fe
grows epitaxially on GaAs  thanks to the small lattice
mismatch (1.4\%). Together with the expected cubic anisotropy of bulk bcc
Fe, a strong in-plane uniaxial anisotropy has been found in ultrathin
Fe/GaAs(001) films, resulting from the atoming bonding at the
interface.\cite{Bland}
The evolution of the latter anisotropy with film thickness has been
quantitatively analyzed by some of us\cite{esteso}
in a thorough {\it in-situ} Brillouin scattering study of the dynamical
magnetic properties of such films.


The same system Fe/GaAs(001) has now been found to display a very
interesting phenomenon, for small iron thickness ($t_{Fe}=6$\AA):
below a critical field $H_c\simeq 0.9$ kOe {\it in-situ} BLS
spectra show a ``double-peak" structure, therefore revealing the
existence of two spin-wave modes for $H<H_c$. This feature is not
completely new: it was already observed\cite{Falco} in Co/Au(111)
films, for $t_{Co} \ge 6$ML and $H < H_c \sim 3$ kOe. However, the
novelty of our contribution is twofold. (i)~The experimental
observation of the double-peak structure has been done in two
different samples and both upon increasing and decreasing  the
magnetic field. This confirms that the phenomenon can be well
reproduced and rules out the possibility it may be due to
metastability effects. (ii)~We develop a theory which explains the
field dependence of the observed spin-wave frequencies as well as
the intensities of the corresponding BLS peaks.

The starting point of our theory is that the observed splitting of
the spin-wave spectrum into two modes is not compatible with a
homogeneous ground state. For this kind of system, the simplest
and most natural explanation is to assume a perpendicularly
magnetized up/down domain structure (for such low values of film
thickness, in-plane magnetized domains are {\it not} energetically
favoured\cite{rassegna,EPJB,Pescia}). Another possibility, a two
sublattices spin arrangement, can be readily disregarded because
the splitting is only observable below a critical value of the
field and because it is hardly applicable to an epitaxial iron
film. Our assumption is therefore an up/down domain ground state.
With increasing the magnetic field $H$, the magnetization of each
domain gradually tilts in the plane and, for $H$ greater than a
critical value $H_c$, the ground state is homogeneously magnetized
in plane. Two branches are found for $H<H_{c}$ and a single
(uniform) mode for $H \ge H_{c}$.

At present an {\it in-situ} high resolution mapping of the
magnetization is outside the reach of conventional microscopic 
techniques. In the absence of detailed
information on the actual domain structure and for the sake of
generality, we are going to assume a stripe domains structure.
Such a one-dimensional model has the advantage that the
frequencies of the spin-wave excitations can be easily evaluated
and their field dependence can be reproduced for different
in-plane directions.

The format of the paper is as follows. In Section II the
experimental details concerning the sample preparation and the BLS
technique are summarized. In Section III the frequencies of the
spin-wave excitations with respect to a non-homogeneous ground
state with up/down stripe domains are calculated using the
Landau-Lifshitz equations of motion. We also estimate the field
dependence of the spin-wave intensity of the two modes in the
framework of a classical macroscopic model which relates the
scattered intensity to the magnetization-dependent permittivity
tensor. In Section IV the experimental results are presented and
compared with the predictions of the theoretical model. Finally,
in Section V the conclusions are drawn. Details about the
calculation of the spin-wave frequencies can be found in Appendix
A.

\section{Experiment}

\begin{figure}
\label{fig1}
\includegraphics[width=8cm,angle=0,bbllx=40pt,bblly=60pt,%
bburx=500pt,bbury=760pt,clip=true]{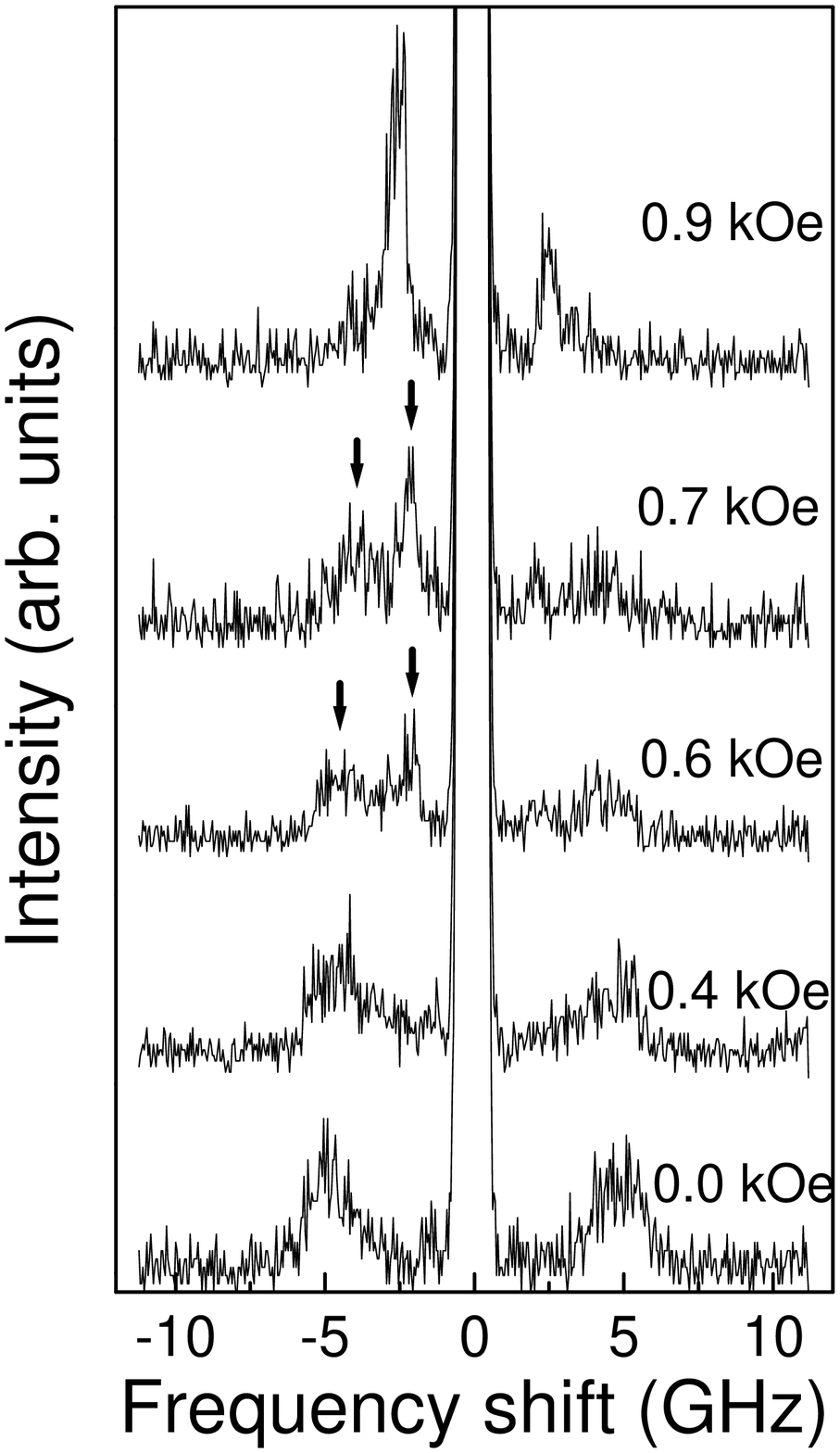}
\caption{
{\it In-situ} Brillouin spectra taken from a $t_{Fe}=6$\AA$~$thick
epitaxial Fe/GaAs(001) film for different values of the external magnetic
field $H$, applied within the film plane along $[-110]$, the hard
in-plane direction. Two peaks, indicated by arrows, are simultaneously
observed for $H=0.6$ and 0.7kOe.
}
\end{figure}

Following the previous investigation of Fe/GaAs(100) films with
different thickness,\cite{esteso} in this work we focus our
attention on 6\AA$~$ thick Fe films, because for this particular
thickness BLS spectra  exhibit the double-peak structure described
in the following. The reproducibility of the results was checked
studying two different samples with the same nominal thickness.
The two iron films were grown on a Si-doped GaAs(001) single
crystal in an ultra-high-vacuum (UHV) chamber, specially designed
to allow {\it in-situ} BLS measurements\cite{apparatus,
apparatus2} (background pressure $3 \times 10^{-10}$ mbar) at
GHOST laboratory, Perugia University.\cite{ghost} Details about
sample preparation and structural characterization can be found in
Ref.~ \onlinecite{esteso}. About 200 mW of monochromatic
p-polarized light, from a solid state laser (532 nm line), was
focused onto the sample surface using a camera objective of
numerical aperture 2 and focal length 50 mm. The back-scattered
light was analyzed by a Sandercock type (3+3)-pass tandem
Fabry-P\'erot interferometer.\cite{Sander} The external dc
magnetic field was applied parallel to the surface of the film and
perpendicular to the plane of incidence of light {\it i.e.} in the
so-called Damon-Eshbach geometry. BLS measurements of the
spin-wave frequency were performed {\it in situ} at room
temperature as a function of both the intensity and the in-plane
direction of the applied magnetic field. Typical BLS spectra for
such a film, taken with the magnetic field applied along the hard
in-plane direction [-110], are shown in Fig.~1. The presence of a
double-peak structure is evident for applied field values
$0.4<H<0.9$ kOe. In contrast, a single peak was observed in the
BLS spectra measured for magnetic field applied along both the
[100] and the [110] directions (not shown).

\section{Theory}

As discussed in the Introduction, the observed splitting of the
spin-wave spectrum into two modes is not compatible with a
homogeneous ground state. In the absence of experimental data
about the actual spin configuration, we assume a simplified model
with perpendicularly magnetized up/down domains in the shape of
parallel stripes of infinite length along the in-plane field
direction and with vanishing thickness of the wall between
opposite domains: see Fig. 2. While more complicated up/down (or
possibly canted) domain patterns cannot in principle be ruled out,
the former one-dimensional model for the non-homogeneous ground
state presents the advantage that the frequencies of the spin-wave
excitations can be easily evaluated using the theory of domain
mode ferromagnetic resonance
(DMFMR),\cite{RW,Ramesh,Ebels,notaBLS} and analytical results can
be obtained for field applied in plane along high symmetry
directions. Such a simple model turns out to be a useful tool to
reproduce the spin-wave behavior; however, one should not expect
it to be entirely realistic or able to account for other
properties of the system, like the domain wall structure, the
spatial dependence of the demagnetization factors, as well as the
domain width and its field dependence.

\begin{figure}
\label{draw}
\includegraphics[width=8cm,angle=90]{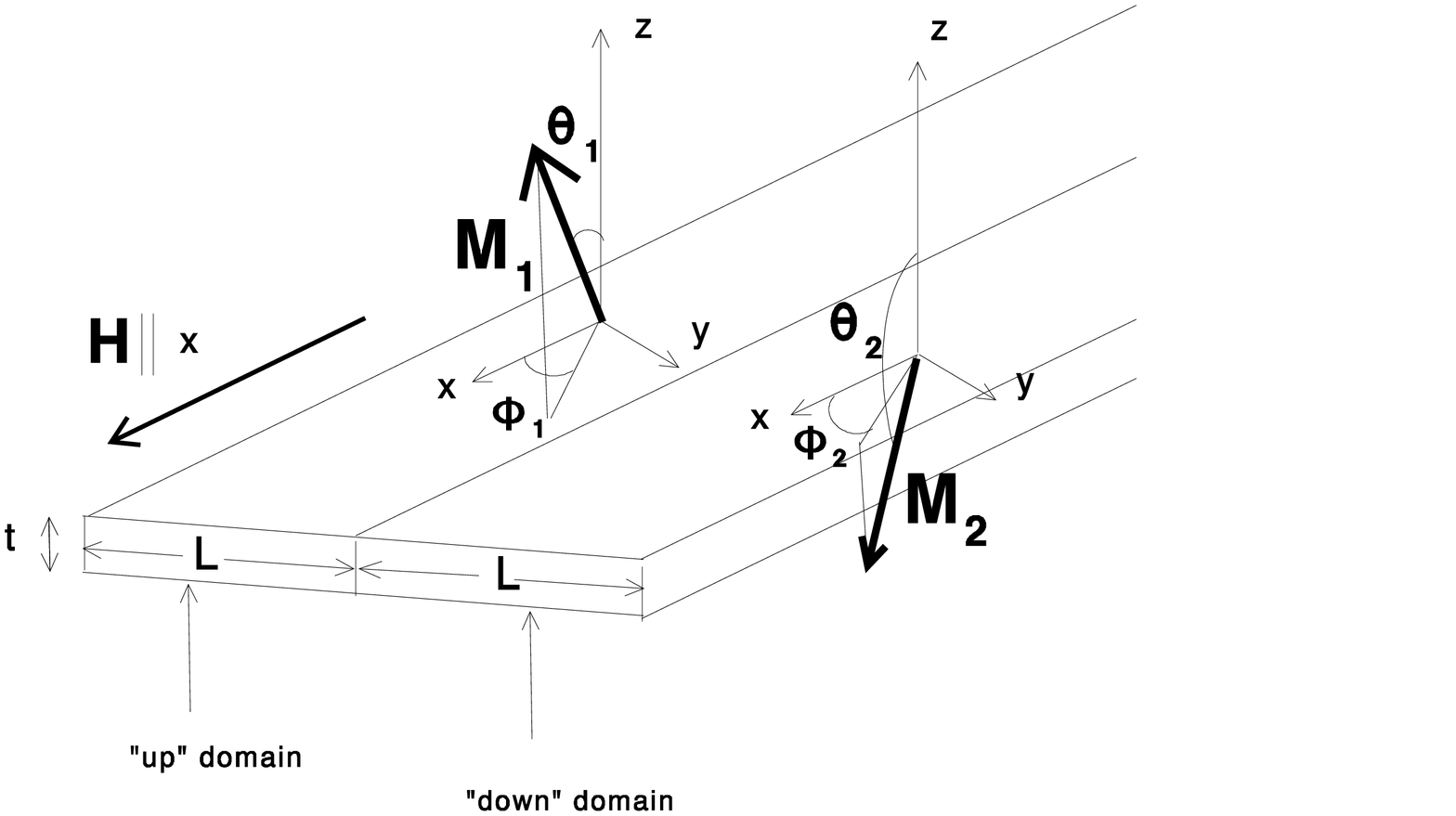}
\caption{One-dimensional up/down stripe domain structure
and coordinate system used to define the domain variables:
$z$ denotes the normal to the film plane.}
\end{figure}

The free energy per unit volume of the system in Fig.~2
is given by the sum of various contributions
\begin{equation}
\label{G}
G=G_{H}+G_{dip}+G_{2\perp}+G_{2\Vert}
\end{equation}
where $G_H$ is the Zeeman energy term due to the external field,
$G_{dip}$ is the term due to demagnetization, while $G_{2\perp}$
and $G_{2\Vert}$ are the terms due to uniaxial out-of-plane and
in-plane anisotropies, respectively. One has\cite{RW}
\begin{eqnarray}
\label{free}
G_H&=&-{{HM_s}\over 2} [\sin\theta_1 \cos\phi_1
+\sin\theta_2\cos\phi_2]
\cr
G_{dip}&=&{{\pi}\over 2}M_s^2[\cos\theta_1+\cos\theta_2]^2
\cr
&+&{{\pi}\over 2}M_s^2 N_{zz}[\cos\theta_1-\cos\theta_2]^2
\cr
&+&{{\pi}\over 2}M_s^2 N_{yy}[\sin\theta_1 \sin\phi_1-\sin\theta_2 \sin\phi_2]^2
\cr
G_{2\perp}&=&-{{K_{2\perp}}\over 2}[\cos^2 \theta_1+\cos^2\theta_2]
\end{eqnarray}
where $M_s$ is the saturation magnetization;
$H$ is the external field applied within the film plane ($xy$)
along the $x$ direction;
$K_{2\perp}>0$ is a uniaxial out-of-plane anisotropy
that favours the up/down spin alignment along $z$,
the normal to the film plane;
$N_{zz}$ is the demagnetization factor associated to the up/down
domain structure: since the stripes are assumed to be parallel to the field direction
($x$ axis), one has $N_{xx}=0$ and $N_{yy}=1-N_{zz}$.
In general, the static demagnetization factor
$N_{zz}$ is a function of the domain aspect ratio $L/t$ (where $L$ is the domain
width and $t$ the film thickness) and of the rotational permeability
$\mu$.\cite{RW,Ramesh,KooyEnz}
For magnetic field applied along the hard in-plane direction,
the in-plane anisotropy contribution to the free energy is written as
\begin{equation}
\label{ahard}
G_{2\Vert}^{(i)}=-{{K_{2\Vert}}\over 2}
[\sin^2\theta_1 \sin^2\phi_1+\sin^2\theta_2 \sin^2 \phi_2]
\end{equation}
while for field applied along the easy in-plane direction
\begin{equation}
\label{beasy}
G_{2\Vert}^{(ii)}=-{{K_{2\Vert}}\over 2}
[\sin^2\theta_1 \cos^2\phi_1+\sin^2\theta_2 \cos^2 \phi_2].
\end{equation}
In the former case the uniaxial in-plane anisotropy
$K_{2\Vert}>0$ favours the $y$ direction (perpendicular to the field and to
the stripes), and in the latter case it favours the $x$ direction
(parallel to the field and to the stripes).

In terms of the free energy parameters, we define the out-of-plane anisotropy
field $H_{2\perp}=2K_{2\perp}/M_s$, the in-plane field $H_{2\Vert}=2K_{2\Vert}/M_s$,
and the dipolar field $H_{dip}=4 \pi M_s$.
It is customary to introduce the quality factor $Q=H_{2\perp}/H_{dip}$
as the ratio between the out-of-plane anisotropy field $H_{2\perp}$,
favouring the perpendicular direction ($z$), and the dipolar field $H_{dip}$,
favouring the film plane ($xy$). In the case under study, we have $Q<1$.
Finally, the saturation field is defined as
$H_{sat}=H_{2\perp}-H_{dip}N_{zz}$.\cite{Ramesh,RW}

In the following, the equilibrium values of the polar and azimuthal
angles, obtained by minimizing the free energy Eq.~(\ref{G}),
will be denoted by the suffix ``{\it e}".
The frequencies of the spin-wave excitations
are evaluated\cite{RW,Ramesh} by the Landau-Lifshitz
equations of motion (see Appendix A for details).
Two modes, denoted by $\omega^+$ (acoustic mode)
and $\omega^-$ (optic mode),
are found for $H<H_{c}$ and a single (uniform\cite{Kittel})
mode for $H \ge H_{c}$.
It is worth noticing that, despite their names, neither of
the two modes is fully in-phase or fully out-of-phase:
their peculiar character is disclosed by the analysis of the
eigenvectors associated to the two modes.\cite{Ebels}
In fact, assuming the external magnetic field to be parallel to
the $x$ direction (along which the domains are infinitely long),
one finds that the acoustic mode with frequency $\omega^+$ is
characterized by dynamic fluctuations
such that $m_1^x(t)+m_2^x(t)=0$ and $m_1^y(t)+m_2^y(t) \ne 0$.
This corresponds to an out-of-phase precession of the dynamic moments
${\bf M}_1$ and ${\bf M}_2$ in the direction parallel to the domain wall
and an in-phase precession perpendicular to the domain wall.
In contrast, for the optic mode the precession parallel to
the domain wall is in-phase and the precession
perpendicular to the domain wall is out-of-phase.\cite{Ebels}

Hereafter we give the expressions of the spin-wave
frequencies\cite{limite} when the field is applied in plane along
the hard axis, see Eq.~(\ref{ahard}), or the easy axis, see
Eq.~(\ref{beasy}). In both cases one has
$\phi_{1e}=\phi_{2e}=0$.\cite{nota2}

$\bullet$ Case ({\it i}): $H$ is along the hard in-plane axis.
For $H<H_c=H_{sat}$, the minimum free energy is obtained
for $\theta_{2e}=\pi-\theta_{1e}$
where $\sin \theta_{1e}=H/H_{sat}$. The frequencies of the acoustic
and optic modes are
\begin{eqnarray}
\label{omegapma}
(\omega^+/\gamma)^2&=&
\left\lbrack
H_{sat}^2-H^2
\Big(
1- {{H_{dip}N_{yy}}\over {H_{sat}}}
\Big)
\right\rbrack
\left\lbrack
1-{{H_{2\Vert}}\over {H_{sat}}}
\right\rbrack
\cr
\cr
(\omega^-/\gamma)^2&=& \left\lbrack H_{sat}^2-H^2 \right\rbrack
\left\lbrack 1+ {{  H_{dip}N_{yy}-H_{2\Vert}  }\over {H_{sat}}}
\right\rbrack
\end{eqnarray}
respectively ($\gamma$ is the gyromagnetic factor).
For $H\ge H_c=H_{sat}$, one has $\theta_{1e}=\theta_{2e}=\pi/2$:
the stripe domain structure is wiped out, and the film is
homogeneously in-plane magnetized. The optic mode ($\omega^-$)
disappears and the acoustic one ($\omega^+$)
becomes the saturated, in-plane, uniform\cite{Kittel} mode with frequency
\begin{equation}
\label{omegaua}
(\omega/\gamma )^2=
\left\lbrack H - H_{2\Vert} \right\rbrack
\left\lbrack H - (H_{2\perp}-H_{dip}) \right\rbrack
\end{equation}

$\bullet$ Case ({\it ii}): $H$ is along the easy in-plane axis.
For $H<H_c=H_{sat}-H_{2\Vert}$, the ground state has
$\theta_{2e}=\pi-\theta_{1e}$ where $\sin \theta_{1e}=H/H_c$
and the frequencies of the acoustic
and optic spin-wave excitations are
\begin{eqnarray}
\label{omegapmb}
(\omega^+/\gamma)^2&=&
H_{sat} H_c \cr
&\times& \left\lbrack
1- \Big({{H}\over {H_c}}\Big)^2 \Big(1-{{H_{dip}N_{yy}}\over {H_c}}\Big)
\right\rbrack
\cr
\cr
(\omega^-/\gamma)^2&=& H_{sat} H_c \cr
&\times& \left\lbrack 1- \Big({{H}\over {H_c}}\Big)^2 \right\rbrack
\left\lbrack 1+ {{  H_{dip}N_{yy} }\over {H_{c}}}
\right\rbrack
\end{eqnarray}
For $H\ge H_c=H_{sat}-H_{2\Vert}$, one has $\theta_{1e}=\theta_{2e}=\pi/2$
and the frequency of the uniform mode is
\begin{equation}
\label{omegaub}
(\omega/\gamma )^2=
\left\lbrack H +H_{2\Vert} \right\rbrack
\left\lbrack H  -(H_{2\perp}- H_{dip})+H_{2\Vert} \right\rbrack
\end{equation}

\section{Results and discussion}

\subsection{Spin-wave frequencies}

\begin{figure}
\includegraphics[width=8cm,angle=0,bbllx=140pt,bblly=85pt,%
bburx=481,bbury=720pt,clip=true]{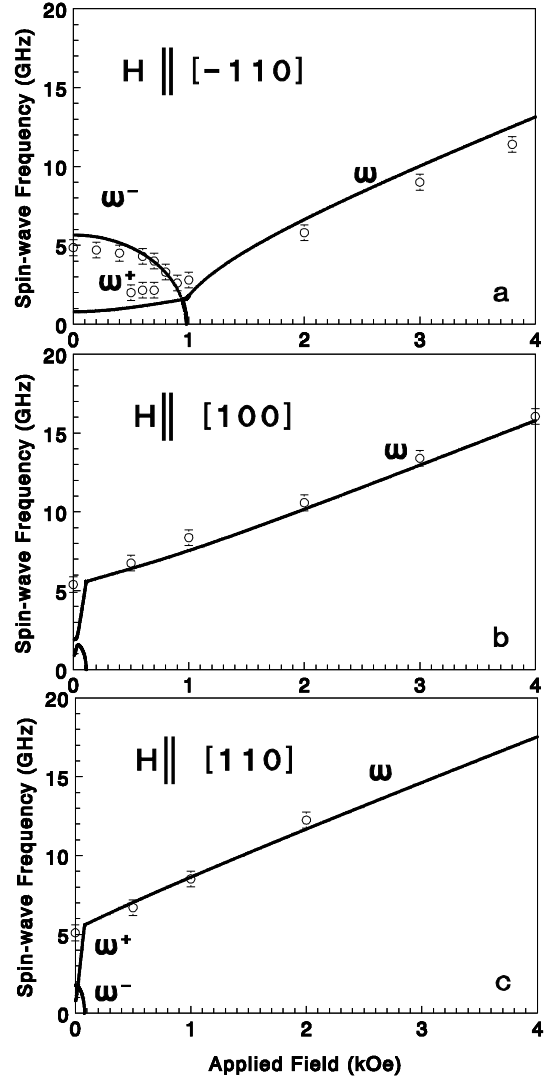}
\caption{Full-line curves: frequencies of the acoustic ($\omega^{+}$),
optic ($\omega^{-}$) and uniform ($\omega$) mode, as a
function of the in-plane field $H$,
calculated using $H_{2\Vert}=0.9$kOe,
$H_{2\perp}=13.9$kOe, $H_{dip}=17$kOe, $N_{zz}=0.76$
($H_{sat}=0.98$kOe), for three different cases:
(a) field applied
parallel to the stripes and along [-110], the hard in-plane
direction (see Eqs. (\protect\ref{omegapma}) and
 (\protect\ref{omegaua}));
(b) field applied parallel to the stripes and
along [100], the intermediate in-plane direction;
(c) field applied
parallel to the stripes and along [110], the easy in-plane
direction (see Eqs. (\protect\ref{omegapmb}) and
 (\protect\ref{omegaub})).
Open circles: experimental data.
}
\end{figure}

In Fig.~3 the measured spin-wave frequencies are plotted as
a function of the intensity of the in-plane applied
magnetic field $H$. When the field is parallel to the hard in-plane
direction [-110] (Fig.~3a), two spin-wave modes are observed
for $0.4<H<0.9$ kOe. No differences in frequency, within experimental error, are
found upon increasing or decreasing the field intensity, thus
ruling out the possibility of metastability effects.
In contrast, a single spin-wave mode is observed for $H \ge 0.9$ kOe.
When the field is applied along either the intermediate [100] or the easy [110]
in-plane directions (see Fig.~3b and 3c, respectively),
a single spin-wave mode is experimentally observed
at any investigated field intensity.
The full-line curves in Fig. 3 are the theoretical spin-wave
frequencies.
The Hamiltonian parameters for the calculations were deduced
from previous experimental work on Fe/GaAs(001) films,\cite{esteso} where
a detailed fit of the BLS data was made for a whole set of samples with
different values of the iron thickness, ranging between $t_{Fe}=4$ \AA$~$
and 100 \AA.  As $t_{Fe}$ increases, the out-of-plane
single-ion anisotropy $K_{2\perp}$ strongly decreases, while the dipolar field
$H_{dip}=4\pi M_s$, favouring in-plane magnetization, increases;
moreover, a biaxial in-plane anisotropy, favouring the [100] and [010]
crystallographic axes, gradually develops.
For the sample with $t_{Fe}=6$ \AA, only the data at fields high enough
for the magnetization to be homogeneous and in plane, were used
to obtain the fit. The dipolar field was estimated to be $H_{dip}=17$kOe,\cite{Trieste}
the out-of-plane anisotropy field $H_{2\perp}=13.9$kOe and the in-plane anisotropy
field $H_{2\Vert}=0.9$kOe.\cite{neglect}
Using these parameters, the demagnetization factor $N_{zz}$ was self-consistently
calculated\cite{KooyEnz} for different values of the domain aspect ratio $L/t$.
The variation of the domain size with the applied field intensity,
and thus the corrections to the frequency due to such variations,
were neglected as being of a higher order.\cite{RW,Ramesh}

The best overall agreement between theory and experiment was found
assuming for the static demagnetization factor the value $N_{zz}=0.76$.
Such an assumption, although it corresponds to a probably too low
aspect ratio,\cite{nota3} is nevertheless able to justify the
presence of domains in the system in spite of the fact that,
for the considered iron thickness $t_{Fe}=6$\AA, one has
$H_{2\perp}<H_{dip}$ (quality factor $Q<1$).
In fact, for the onset of up/down stripe domains,
the condition $H_{2\perp}-H_{dip}N_{zz}>0$ has to be satisfied.
As a further support to the domain hypothesis, it is worth observing that
in epitaxial Co/Pt multilayers a perpendicular (up/down) stripe domain structure
was indeed experimentally observed at remanence by magnetic force microscopy,
while torque magnetometry measurements, providing $Q<1$,
had suggested a preference for
in-plane orientation of the magnetization.\cite{Stamps}

Note that the two modes are well observable only for the case of
field applied along the hard in-plane direction (Fig.~3a).
Otherwise (see Fig.~3b,c) one has a considerable shrinking of the
coexistence region of the two modes and moreover metastability
phenomena are likely to occur since the energy of the stripe
domain ground state is close to that of a homogeneous in-plane
configuration.

\subsection{Light scattering intensities}


For the experimental backscattering geometry
($H$ in plane parallel to the $x$ axis and scattering plane perpendicular to $H$)
the incident light has p-polarization
(the optical wavevector and the optical incident electric field {\bf E}$_I$
have only $y$ and $z$ components)
while the scattered light has s-polarization. Then its intensity is
proportional to the square modulus of the $x$ component of the
polarization {\bf P} induced in the
film\cite{Cottam,Mills,DutcherJMMM,TesiDutcher,Zivieri}
\begin{eqnarray}
\label{tesi}
4\pi P_x&=&m^x(t)~~ \Big\lbrack
-K E^y_I
\sin\theta_{e}\cos\theta_{e}
\cr
&-&2G_{44}M_s E^z_I
\cos\theta_{e}(\sin^2\theta_e-\cos^2\theta_e)
\Big\rbrack
\cr
\cr
&+& m^y(t)~~ \Big\lbrack
2G_{44}M_s E^y_I\sin\theta_e  -KE^z_I
\Big\rbrack
\cr
\cr
&+& m^z(t)~~\Big\lbrack
K E^y_I\sin^2\theta_e
\cr
&+& 2G_{44}M_s E^z_I
\sin\theta_e(\sin^2\theta_e-\cos^2\theta_e)
\Big\rbrack
\end{eqnarray}
\noindent where $K$ and $G_{44}$ denote the first- and second-order
(complex) magneto-optic coupling coefficients, respectively.
For the film with up/down domains, we assume that
$m^{\alpha}(t)=m_1^{\alpha}(t)+m_2^{\alpha}(t)$
($\alpha=x,y,x$) since the size of the laser spot is much greater
than the lateral size of the domains.\cite{Zivieri}
Taking into account that $\phi_{1e}=0$ and $\sin\theta_{1e}=H/H_c=h$
and expressing the dynamic fluctuations of the magnetization in terms of the
fluctuations of the angle coordinates $\Delta \theta^{\pm}(t)$, $\Delta \phi^{\pm}(t)$
defined in Appendix A, we obtain
\begin{eqnarray}
\label{amplitude}
4\pi P_x &=& \Delta \theta^-(t)
~~\Big\lbrack
-KE^y_I ~h (1-h^2)
\cr&-&2G_{44}M_s E^z_I ~(1-h^2)(2h^2-1)
\Big\rbrack
\cr
&+& \Delta \phi^+(t)~~ \Big\lbrack
2G_{44}M_s E^y_I~ h^2  -KE^z_I ~h
\Big\rbrack
\cr
&+&\Delta \theta^+(t)~~\Big\lbrack
-K E^y_I ~ h^3 \cr
&-& 2G_{44}M_s E^z_I ~ h^2 (2h^2-1)
\Big\rbrack
\end{eqnarray}
The field dependence of the intensities of the two modes can now be estimated
taking into account that the eigenvector associated with the acoustic mode
is characterized by $\Delta \theta^-(t)=\Delta \phi^-(t)= 0$
while the optic mode
has $\Delta \theta^+(t)= \Delta \phi^+(t)= 0$.

$\bullet$ Acoustic mode with frequency $\omega^+$.
The intensity $I^+(H)$ of the light scattered by the
acoustic mode has a maximum for $H \to H_c$ since the
fluctuations become very large. For zero field
the intensity is zero, $I^+(0)= 0$,
since $\Delta \theta^-(t)= 0$ and the coefficients
of the ``$^+$" angle fluctuations are zero for $h=0$.
For $H\to + \infty$, the intensity vanishes, $I^+(H)\to 0$, since,
upon increasing $H$ above $H_c$, the $\omega^+$ mode
evolves into the uniform mode $\omega$ and
the fluctuations progressively decrease.
This behavior for $I^+(H)$ is similar to that of a perpendicularly
magnetized uniform film.\cite{Dutcher}

$\bullet$ Optic mode with frequency $\omega^-$.
For $H=H_c$, the intensity of the light scattered by the optic mode
is zero $I^-(H_c)= 0$, since $\Delta \theta^+(t)=
\Delta\phi^+(t)=0$ and the coefficient of the $\Delta \theta^-(t)$
angle fluctuation is zero for $h=1$.
For zero field, the intensity $I^-(0)$ can be finite provided
that the second order magneto-optic coupling coefficient is
nonzero, $G_{44}\ne 0$.

The field dependence of the intensity of both the acoustic and the
optic modes, as deduced from Eq.~(\ref{amplitude}), is
qualitatively confirmed by the experimental spectra in Fig.~1.
The intensity of the former mode exhibits a neat maximum for field
values sligtly lower than $H_{c} \approx 0.9$ kOe and then it vanishes
as the field is reduced below about $0.5$ kOe.
The optic mode intensity, instead, shows a minimum approaching
$H_c$, in agreement with the theoretical predictions.

\section{Conclusions}

In conclusion, we have shown that a double-peaked structure is displayed
by the Brillouin Light Scattering spectra of Fe/GaAs(001) films with
$t_{Fe}=6$\AA$~$ when the field is applied in plane along the hard axis
and is smaller than a critical value $H_c=0.9$ kOe.
The existence of two peaks in the BLS spectrum should be the
general feature of a film with a perpendicular domain
structure (it is irrelevant whether the magnetization is canted or not).
This feature disappears when $H \ge H_c$ and the magnetization lies
in the film plane.
The reason why the unravelling of such a two-peaked structure
in the Brillouin light scattering spectra of ultrathin magnetic
films is so rare might well be that many conditions have to be
simultaneously satisfied.
In fact, the optic mode, with frequency $\omega^-$,
has enough intensity in an appreciable
range of fields only if $G_{44}\ne 0$
and $H_c$ are not too small.
In contrast, the acoustic mode, with frequency $\omega^+$,
has more chances to be observed since its intensity,
though always vanishing in the $H\to 0^+$ limit, is expected
to increase as $H$ increases and to reach a maximum just at $H_c$.
Another stringent requirement for the simultaneous observation
of two modes is that the competing out-of-plane anisotropy field $H_{2\perp}$
and easy-plane dipolar anisotropy field $H_{dip}$ are of comparable
magnitude and that $H_{2\perp}-H_{dip}N_{zz}>0$,
so that a perpendicularly magnetized up/down
domain structure is energetically favoured for $H \to 0^+$.
This seems just to be the case of the Fe/GaAs films
with $t_{Fe}=6$\AA. In fact, for higher Fe thickness, one has
$H_{dip} \gg H_{2\perp}$, so that a homogeneous in-plane magnetized
ground state is realized, while, upon reducing the Fe thickness,
one would expect $H_{dip} \ll H_{2\perp}$ and
a single spin-wave mode to be excited with respect to a homogeneous,
perpendicularly magnetized metastable state.

We hope that the results of this paper can stimulate other
experimental groups to directly visualize the domain pattern, {\it
e.g.} using magnetic  microscopy techniques  as a function of the
external magnetic field intensity.\cite{Meyer} This should be done
{\it in situ}, because the magnetic anisotropy is strongly
affected by the presence of a protective overlayer, so that
formation of magnetic domains can be prevented.\cite{esteso}

\begin{acknowledgments}
The authors gratefully acknowledge financial support from the Italian
Ministery for the Instruction, University and Research (MIUR), under projects
PRIN 2003025857 and FIRB RBNE017XSW. This work was performed in the
framework of the joint CNR-MIUR programme (Legge 16/10/2000, Fondo FISR).
\end{acknowledgments}

\appendix
\section{Spin-wave frequencies}

The static equilibrium configuration of the system is obtained
by minimizing the free energy $G$, Eq.~(\ref{G}), with respect
to the polar and azimuthal variables
while the frequencies of the spin-wave excitations
are evaluated \cite{RW} by the Landau-Lifshitz
equations of motion
\begin{equation}
{{d \theta_{1,2}}\over {dt}}= - {\gamma\over {\sin\theta_{1,2}}}
{{\partial {\cal G}}\over {\partial \phi_{1,2}}},
~~~~
{{d \phi_{1,2}}\over {dt}}=  {\gamma\over {\sin\theta_{1,2}}}
{{\partial {\cal G}}\over {\partial \theta_{1,2}}}
\end{equation}
where ${\cal G}=2G/M_s$. The small oscillations of the
system in response to an external perturbation
are obtained by expanding ${\cal G}$ in a Taylor series about its equilibrium value
${\cal G}_e$ up to the second order.
Next, assuming for the set of variables
a harmonic time dependence with frequency $\omega$ and
introducing the normal coordinates $\Delta \theta^{\pm}=\Delta \theta_1 \pm \Delta
\theta_2$ and $\Delta \phi^{\pm}=\Delta \phi_1 \pm \Delta \phi_2$
(where $\Delta$ denotes a small variation), the
equations of motion can be rewritten in matrix form as
\begin{equation}
\label{matricial}
\left\lbrack
\begin{array}{l}
        ~A^+~ \; -iz \; ~~0~ \; ~~B^+ \;
        \cr
        ~iz \; ~~~~~C^+~ \; B^-~\; 0~\;
        \cr
        ~~0~~ \; ~~~B^-~\; A^-\; -i z\;
        \cr
        ~B^+~ \; ~~0\;~~~~i z \;~~~ C^-
\end{array}
\right\rbrack
\left\lbrack
\begin{array}{l}
        \Delta \theta^+ \;
        \cr
        \Delta \phi^+
        \cr
        \Delta \theta^- \;
        \cr
        \Delta \phi^- \;
\end{array}
\right\rbrack =0
\end{equation}
where $z=\omega ~\sin \theta_{1e}$ and $A^{\pm}={\cal G}_{11} \pm
{\cal G}_{13}$, $B^{\pm}={\cal G}_{12} \pm {\cal G}_{23}$,
$C^{\pm}={\cal G}_{22} \pm {\cal G}_{24} $. By ${\cal G}_{ij} = {
{\partial^2 {\cal G} }\over {\partial X_i \partial X_j}
}\Big\vert_e$ we denote the second partial derivatives of the free
energy with respect to the angular variables
($X_1=\theta_1,X_2=\phi_1,X_3=\theta_2,X_4=\phi_2$). The
frequencies of the normal modes are obtained by imposing the
condition for nontriviality of the solutions of Eq.
(\ref{matricial}), {\it i.e.} the vanishing of the matrix
determinant.

For $H \ge H_c$ the ground state is homogeneously in-plane
magnetized, ($\theta_{1e}=\theta_{2e}=\pi/2$ and $N_{zz}=1$) and
it results that $A^+=A^-={\cal G}_{11}$, $B^+=B^-=0$, and
$C^+=C^-={\cal G}_{22}$, so that a single, uniform
mode\cite{Kittel} is obtained, with frequency $\big( \omega/\gamma
\big)^2 = {\cal G}_{11} {\cal G}_{22}$.


\begin{thebibliography}{99}
\bibitem{limite}
It is worth noticing that in the absence of in-plane anisotropy, $H_{2\Vert} \to 0$,
the results of Ref. \onlinecite{RW} for the mode frequencies (Eqs.~31, 32, 35)
are correctly recovered.
\bibitem{nota2}
This is obvious for field applied along the
easy axis. For field along the hard axis,
this is verified provided that $H_{2\Vert}<H_{sat}$,
as is the case for the system under study;
when $H_{2\Vert}>H_{sat}$,
the excitation of two modes would not be allowed since the ground
state is characterized by a homogeneous magnetization lying within
the film plane.
\bibitem{neglect}
For the sake of simplicity we neglect the
anisotropy field $H_{4\Vert} \approx 0.1$kOe associated to a biaxial in-plane
anisotropy that favours the [110] and [-110] axes, since it is much smaller than
the uniaxial in-plane anisotropy fied $H_{2\Vert}$.\cite{esteso}
\end{thebibliography}
\end{document}